\documentclass[12pt,english]{article}
\usepackage[T1]{fontenc}
\usepackage[latin9]{inputenc}
\usepackage{geometry}
\usepackage{color}
\usepackage{babel}
\usepackage{float}
\usepackage{graphicx}
\usepackage{setspace}
\usepackage[unicode=true,pdfusetitle,
 bookmarks=true,bookmarksnumbered=false,bookmarksopen=false,
 breaklinks=false,pdfborder={0 0 0},backref=false,colorlinks=true]
 {hyperref}

\makeatletter
\usepackage{enumitem}		

\@ifundefined{date}{}{\date{}}
\usepackage{lmodern}
\usepackage[T1]{fontenc}
\usepackage{caption}
\usepackage{upgreek}

\@ifundefined{showcaptionsetup}{}{%
 \PassOptionsToPackage{caption=false}{subfig}}
\usepackage{subfig}
\makeatother

\begin{document}

\title{Co-axial Helicity Injection on the STOR-M Tokamak}

\author{Carl Dunlea$^{1*}$, Chijin Xiao$^{1}$, and Akira Hirose$^{1}$}

\maketitle
$^{1}$University of Saskatchewan, Saskatoon, Canada 

$^{*}$e-mail: cpd716@mail.usask.ca 
\begin{abstract}
Injection of relatively high density spheromaks with significant helicity-content
into a tokamak has been proposed as a means for fueling and current
drive. The CHI (Co-axial Helicity Injection) device was devised to
inject current to the STOR-M tokamak. Various circuit modifications
were made to the CHI controls, enabling testing of various injection
configurations. The charge/discharge circuits for CT formation/acceleration
and stuffing field were modified, and the power supplies and power
converters were replaced. Various modifications were implemented to
solve the original slow bank triggering problems. The CHI device was
mounted on STOR-M for radial and vertical CT injection at various
times. Spheromak injection into STOR-M usually resulted in disruption
of the tokamak discharge. After modifying the CHI device to operate
at increased power, it looked like tokamak current was increased by
a few kiloamps just prior to disruption, but careful testing proved
that the signals indicating a current drive were actually spurious,
caused by inductive pickup. The CHI device was attached to a portable
vacuum chamber that was constructed from spare parts, to characterise
the CTs produced. Magnetic probes were constructed to measure poloidal
and toroidal field near the CT edge. Langmuir probes were made and
returned reasonable estimates for edge CT density and temperature. 
\end{abstract}

\section{Introduction}

Injection of relatively high density spheromaks with significant helicity-content
into a tokamak has been proposed as a means for fueling and current
drive. In helicity injection current drive, helicity conserving magnetic
relaxation processes incorporate helicity added to the system by dissipating
any excess free energy, increasing parallel currents and keeping the
plasma close to the Taylor state \cite{REdd}. The spheromak's field
reconnects (relaxes) to that of the tokamak to add helicity. It is
because helicity is conserved even in the presence of turbulent tearing
that current can be expected to be driven as a result of helicity
injection \cite{4brown}. For successful helicity injection, spheromak
resistive decay must not conclude before it has a chance to travel
from its formation region to the tokamak core. Helicity is injected
by driving current on injector magnetic flux, $\Psi_{main}$. The
helicity injection rate is given by $2V_{gun}\Psi_{main}$ \cite{Brown_Bellan,JArboe_review_spheromak},
where $V_{gun}$ is the voltage across the helicity injection device
electrodes, and $\Psi_{main}$ is the injector stuffing flux. The
USCTI (University of Saskatchewan Compact Torus Injector) has demonstrated
core plasma density increase and momentum injection on the STOR-M
tokamak \cite{USCTI_XIAO,USCTI_AKBAR}. The CHI (Co-axial Helicity
Injector) device was devised to inject current to STOR-M. As described
in \cite{Asai1}, a similar injector was used to inject helicity to
the TPE-RX reversed field pinch. The main difference between the USCTI
and CHI devices are that the USCTI works at higher power - it has
a gun voltage of tens of kilovolts with operation over a few microseconds
- whereas the CHI device was originally designed to operate at a few
hundred volts over several milliseconds. Helicity injection (current
drive) has not been observed with the USCTI. \\
\\

\section{\label{sec:Operation-of-the}Operation of the MCPG}

The magnetised co-axial plasma gun (MCPG) used in this project was
designed by Professor Asai at Nihon University, and several visiting
students at the University of Saskatchewan plasma physics laboratory
have worked on getting it to produce plasma. 
\begin{figure}[H]
\centering{}\includegraphics[scale=0.35]{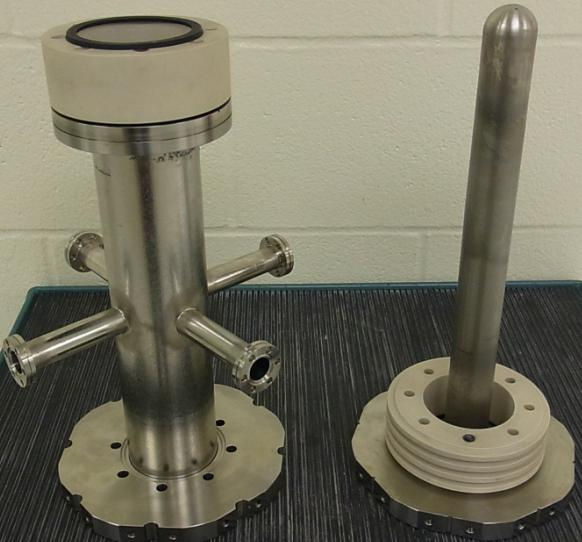}\caption{\label{fig:3}Main CHI components before assembly. }
\end{figure}
 Figure \ref{fig:3} shows a photograph of the inner and outer electrodes
before assembly. One of the four radially oriented access ports has
been blanked off; two are used to inject gas as a plasma source and
one is vacuum sealed with a quartz window for optical diagnostics.
Both electrodes are made of solid stainless steel. In the original
design, the inner electrode is the anode. The two white hollow cylinders
are made from PEEK (polyether ether ketone) and provide electrical
insulation between the anode and cathode and between the cathode and
the vessel (tokamak or test chamber) into which the CHI-produced plasma
is injected. The black viton gasket visible resting on the top of
the cathode's PEEK insulator is one of four such gaskets used for
vacuum seal between the insulators and the electrodes and vessel.
These gaskets are not heat-resistant and their proximity to the CHI-produced
plasma led to melting during testing in 2011. Further occurrence of
this event was prevented by inserting two alumina sleeves (thin-walled
hollow cylinders) inside the PEEK insulators - shielding the gaskets
from the hot plasma.

The working gas used was generally hydrogen, which was injected into
the vacuum-space between the electrodes. The amount of gas required
for plasma formation depends on the strength of the stuffing field
and on the gun voltage. It is thought to be optimal to apply the gun
voltage just before or coinciding with the end of the gas puff. If
the voltage is applied later, the gas will have the opportunity to
diffuse away from the gas-injection region, leading to an increase
in the concentration of neutral gas entrained with the plasmoid through
collisions with accelerating plasma ions. The main solenoid itself
is comprised of 30 turns of 10-gauge wire - 10 turns are wound on
the cathode in the region below the access ports and a further 20
turns are wound (two layers) above the ports. The usual rise time
of the solenoid current is around 1ms, so the solenoid bank can be
discharged to allow for peak stuffing field coincident with actual
plasma breakdown.

\section{CHI charge/discharge circuits}

\subsection{\label{sub:CHI-gun-current}CHI gun current circuit}

\begin{figure}[H]
\begin{centering}
\includegraphics[scale=0.77]{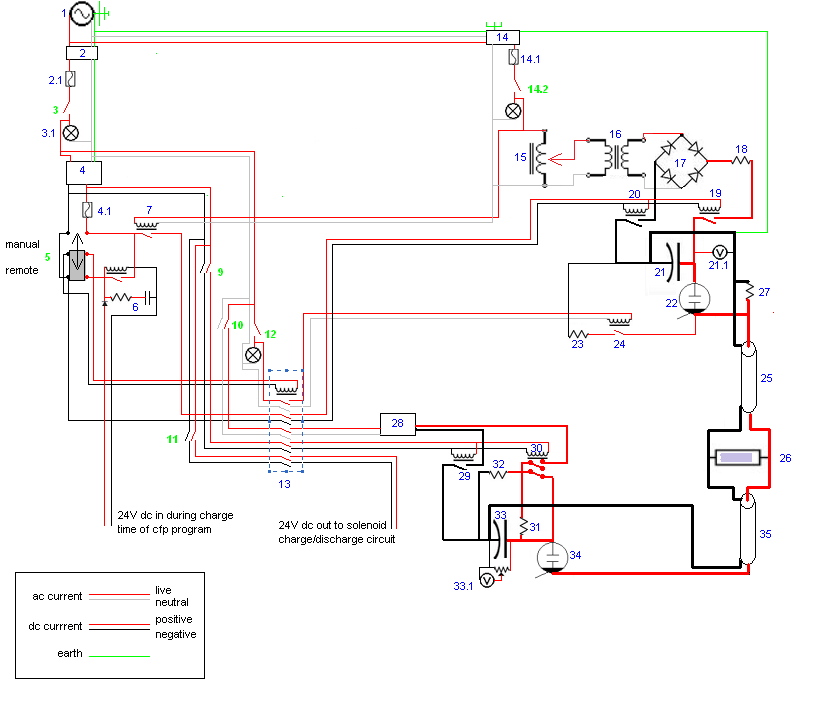}
\par\end{centering}

\caption{\label{fig:9-1}CHI fast \& slow bank charge/discharge circuit}
\end{figure}
The original circuit for the charging and discharging of the main
CHI capacitor banks is shown in figure \ref{fig:9-1}. Current from
the 115V ac supply (1) is split into two main branches and is filtered
at (2) and (14) to remove irregularly high voltage spikes that might
damage other circuit components. (3) is the control circuit power
switch. (2.1) is a 2 Amp fuse, (3.1) is a power indicator lamp. (4)
is a 115V ac to 24V dc power converter - the control system works
with 24V dc. Switch (5) allows operation in manual mode or remote
mode. In manual mode, 24V dc current can flow to the solenoids of
control relays \#3, \#4 and \#5 (depicted here, for simplicity, as
the single solenoid in (13)). Energisation of these three solenoids
causes the five switch-pairs in (13) to close. In remote mode, the
24 V dc signal generated during the charge time of the cfp (control,
fire, position) STOR-M cycle energises the solenoid of the relay in
sub-circuit (6), closing the five switch-pairs in (13).

Switch (10) is the fast bank power switch. When (10) is closed (fast
bank \textquotedbl{}on\textquotedbl{}) and the relays in (13) are
closed, ac current can flow to the 12mA 115V ac to 0 - 5kV (variable)
dc power convertor at (28). Switch (9) is the fast bank charge/dump
switch. In \textquotedbl{}charge\textquotedbl{} mode, switch (9) is
closed and (if the relays in (13) are closed) dc current can flow
to close the fast bank charge relays (29) and (30), so that high voltage
dc current from (28) can flow through the fast bank charge resistor
at (31) to charge the fast capacitor at (33). The original fast capacitor
bank was a 2$\upmu$F unit (two 4$\upmu$F capacitors in series) with
$2\mbox{W},\:265\mbox{k}\Omega$ charge resistor. Later this was modified
to a $0.5$mF capacitor with the same charge resistor. The fast and
slow bank charge resistors are there to prevent damage to the components
in the charging sources by limiting the currents drawn from them.
A voltmeter connected across a 1000:1 voltage divider for monitoring
the voltage on the capacitor is depicted at (33.1). In \textquotedbl{}dump\textquotedbl{}
mode, switch (9) is opened, opening the relays (29) and (30). This
prevents charge current from flowing and allows the capacitor to dump
its energy through the fast bank dump resistor at (32). (32) was originally
a $2\mbox{W},\:10\mbox{M}\Omega$ resistor and was changed to $100\mbox{W},\:5\mbox{k}\Omega$
to handle the longer dump times associated with the $0.5$mF capacitor.
The fast and slow bank dump resistors are there for safety - they
dissipate power in the event that either or both of the ignitrons
fail to fire.

Switch (14.2) is the slow bank power switch. When it is closed (slow
bank \textquotedbl{}on\textquotedbl{}), ac current flows to the variac
(15) and control relay \#2 (7). This current flows through the 15A
slow-blow fuse ((14.1) - prevents slow bank charge current above the
maximum VA rating of the transformer at (16) from being drawn) and
through the the \textquotedbl{}slow bank power on\textquotedbl{} indicator
lamp. When (7) is closed and the relays in (13) are closed, dc current
can flow to close the slow bank charge relays (19) and (20), completing
the slow bank charge circuit. The output of the variac is ac current
at between 0 and 130Volts. This current is stepped up in voltage at
a 1:5.8 transformer (16). The original bridge rectifier (17) was replaced
with a module comprising four A6F120 stud diodes to handle the extra
charge-current drawn by the modified slow bank. The output of the
rectifier bridge is a pulsating dc current at between 0 and 800 Volts
(limited to prevent over-charging of the slow bank).

Switch (12) is the slow bank charge/dump switch. When it is closed
(\textquotedbl{}charge\textquotedbl{} mode) and the relays in (13)
are closed, the slow bank dump relay (24) is opened, preventing dumping
of slow bank charge through the slow bank dump resistor ((23) - originally
175W, $50\Omega$, changed to 175W, $150\Omega$ with the slow capacitor
bank modification). With switches (3), (12) and (14.2) set to \textquotedbl{}on\textquotedbl{},
\textquotedbl{}charge\textquotedbl{} and \textquotedbl{}on\textquotedbl{}
respectively, with the relays at (13) closed (ie. (5) is set to \textquotedbl{}manual\textquotedbl{}
or is set to \textquotedbl{}remote\textquotedbl{} and control relay
\#1 in (6) is closed by the 24V dc current present during the charge
section of the cfp cycle), dc current can flow from the rectifier
bridge through the slow bank charge resistor (18) to charge the slow
capacitor bank (21). (18) was originally 225W, $50\Omega$, changed
to 225W, $150\Omega$ with the slow capacitor bank modification. A
voltmeter (21.1) monitors voltage across the slow bank.

STOR-M's cfp cycle is composed of three stages - charge, fire and
position. During the \textquotedbl{}charge\textquotedbl{} stage, 24V
dc current flows to various relays that switch to allow charging of
the various capacitor banks associated with STOR-M operation. If the
CHI device is selected to be part of the cfp cycle, relay \#1 (in
(6)) and relays \#3, \#4 and \#5 (all in (13)) are closed (with switch
(5) set to \textquotedbl{}remote\textquotedbl{}) so the CHI fast and
slow capacitor banks can be charged. In this case, switches (3) (control
power), (5) (operation mode), (9) (fast bank control), (10) (fast
bank power), (12) (slow bank control) and (14.2) (slow bank power)
are set to \textquotedbl{}on\textquotedbl{}, \textquotedbl{}remote\textquotedbl{},
\textquotedbl{}charge\textquotedbl{}, \textquotedbl{}on\textquotedbl{},
\textquotedbl{}charge\textquotedbl{} and \textquotedbl{}on\textquotedbl{}
respectively. If charging of the CHI solenoid capacitor bank is required,
switch (11) should be set to \textquotedbl{}on\textquotedbl{}. The
\textquotedbl{}fire\textquotedbl{} signal involves termination of
the 24V dc signal in and pulses along the optical fibers to four individual
6V trigger circuits (see sub-section \ref{sub:Optically-triggered-6V}).
The optically triggered 6V trigger circuits generate 6V pulses. Two
of these four 6V pulses trigger the formation of 1.4kV pulses (see
section \ref{sub:Trigger-circuit-for}) to the ignitors of the ignitrons
((34) and (22)), allowing discharge of the fast and slow capacitor
banks through the co-axial transmission lines ((35) and (25)) to the
CHI device electrodes (26). If the ignitrons have fired but conditions
are such that capacitor discharge through a CHI plasma is prevented,
then both capacitor banks can discharge through the \textquotedbl{}dummy\textquotedbl{}
resistor (27). This was originally 100W, $10\Omega$ - changed to
100W, $30\Omega$ with the slow bank capacitor modification. 

The third 6V pulse switches the SCR ((1) in figure \ref{fig:7}) to
allow discharge of the solenoid capacitor bank. The fourth 6V pulse
has a duration that is controlled by an input to the cfp program.
The other 3 6V pulses are all set at around 0.5 ms by the 6V trigger
circuits. The duration of the fourth 6V pulse is the duration for
which current (driven by a variable 0-100V dc source) is allowed to
flow (via co-axial cable) to open the CHI gas valve.

Sub-circuit (6) in figure \ref{fig:9-1} above is made up of control
relay \#1 with a combination of a diode, 500$\Omega$ resistor and
a 2mF capacitor. The RC time for the resistor/capacitor combination
is such that current will flow to keep the relay closed for around
five seconds after the \textquotedbl{}fire\textquotedbl{} signal (termination
of the 24V dc in) from the cfp system. The relay is kept closed for
an additional few seconds after the fire signals in order to prevent
the charged capacitor banks from immediately dumping through the fast
and slow dump resistors ((32) and (23)). Depending on the amount of
neutral gas injected, it can take up to several tens of ms for the
neutral gas to break down and for the capacitor bank-driven current
to start to flow through the plasma. The diode in (6) forces the capacitor-driven
current to flow through the relay solenoid. Sub-circuit (6) prevents
dumping of the energy before it is given a chance to be useful ($i.e.,$
through plasmoid formation). 

\begin{figure}[H]
\centering{}\includegraphics[scale=0.5]{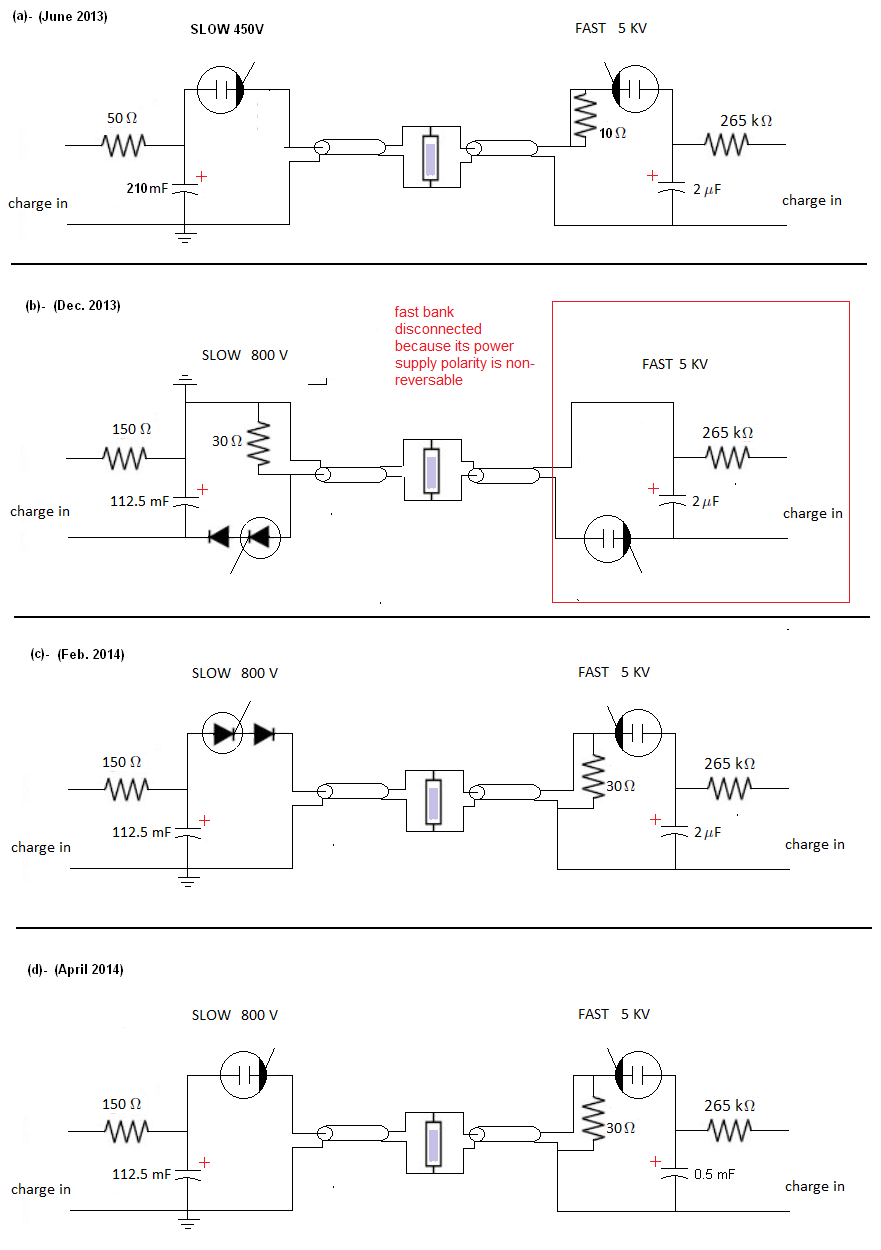}\caption{\label{fig:Various-gun-current}Various gun current circuit configurations }
\end{figure}
Figure \ref{fig:Various-gun-current} indicates four of the fast \&
slow bank charge/discharge circuit configurations used. For simplicity
the control and dump part of the circuits are omitted. The original
circuit (figure \ref{fig:Various-gun-current}(a)) was modified first
(aside from general re-wiring improvements) by extending the slow
bank as described in section \ref{sub:Slow-capacitor-bank.}. This
extension also required replacing wires and connectors with ones that
are able to handle the higher current loads during capacitor discharge.
Also, the slow bank ignitron was replaced with an SCR (model T9G0121203DH;
$V_{RRM}$ (maximum repetitive peak reverse voltage rating) = 1.2kV,
$I_{FSM}$ (Max. forward surge current) =27 kA) and a stack of three
puck diodes (model R9G022415ASOO; $V_{RRM}$ = 2.4kV (so 7.2kV for
the stack), $I_{FSM}$ =15kA) to fix the problem with the slow bank
ignitron trigger circuit (a substantial part of the fast bank voltage
would appear across the slow bank trigger output causing failure of
the SCR in that circuit - see section \ref{sub:Trigger-circuit-for}).
The SCR (silicon controlled rectifier) here is triggered directly
by the optically triggered 6V pulse (section \ref{sub:Optically-triggered-6V}).
The diode stack functions to prevent failure of the SCR by blocking
the fast bank signal. Later, as depicted in figure \ref{fig:Various-gun-current}(b),
the slow bank circuit was modified to allow for a negatively biased
central CHI device electrode, with the positive terminal of the slow
bank connected to ground. The polarity of the fast bank power convertor
(for fast bank charging) is non-reversible so the fast bank was disconnected
in this configuration. The motivation for the polarity change was
that the results of one co-axial helicity injection experiment \cite{4brown}
indicated that tungsten or chromium plating on the CHI cathode was
required for helicity injection, and the inner stainless steel electrode
of the CHI device here had previously been chromium-plated. The theory
is that sputtering at the steel cathode from the impact of positive
ions can add to the impurity content of the injected plasmoid, resulting
in disruption of the tokamak plasma. However, as described in section
\ref{sec:Results-of-CHI}, disruption was usual even with this modification.
Later, when the CHI device was mounted radially on the tokamak, the
electrode polarity was reversed again as shown in figure \ref{fig:Various-gun-current}(c).
Figure \ref{fig:Various-gun-current}(d) indicates the most recent
configuration, with a 5mF fast bank and the SCR and diode stack replaced
with an ignitron again. As described in section \ref{sub:Trigger-circuit-for},
a set of metal oxide varistors was used to fix the slow bank trigger
problem after the SCR (in figure \ref{fig:Various-gun-current}(c))
failed.
\begin{figure}[H]
\subfloat[]{\includegraphics[width=7.5cm,height=6.5cm]{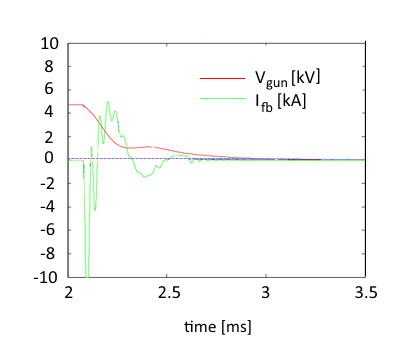}}\hfill{}\subfloat[]{\includegraphics[width=7cm,height=5cm]{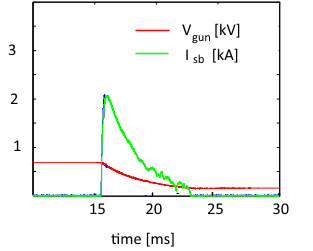}}\caption{CHI voltage and current for fast and slow banks\label{fig:CHI-voltage-and} }
\end{figure}
Figure \ref{fig:CHI-voltage-and}(a) shows the CHI device gun voltage
and gun current for a discharge using only the 5kV, 0.5mF fast bank.
Note that the gun current is oscillatory, exhibiting an RLC discharge.
Figure \ref{fig:CHI-voltage-and}(b) shows the CHI device gun voltage
and gun current for a discharge with 600V on bank1 (50mF) of the slow
bank. Note that the gun current is that of an RC discharge - the safety
diodes on the slow bank prevent an oscillatory discharge. The gun
voltage signals were measured using a 1000:1 probe. The current signal
in figure \ref{fig:CHI-voltage-and}(b) is from a hand-wound Rogowski
coil (with a 23ms RC integrator) that was calibrated against a commercial
iron-cored coil (Pearson model 101). The Pearson Rogowski has a built-in
integrator. For long current pulses like this one, its signal is useful
only for measuring the peak - at later times the signal is not meaningful
as the iron core has become saturated. The current signal from the
hand-wound Rogowski was also calibrated against the voltage drop measured
by differential probe across a known resistance that could handle
the load- 10 feet of 8-gauge wire in this case. The solenoid current
measurements (not shown here) are from the voltage drop measured by
differential probe across $R_{ss}$, the solenoid series resistance.

\subsubsection{\label{sub:Slow-capacitor-bank.}Slow capacitor bank}

\begin{figure}[H]
\centering{}\includegraphics[scale=0.5]{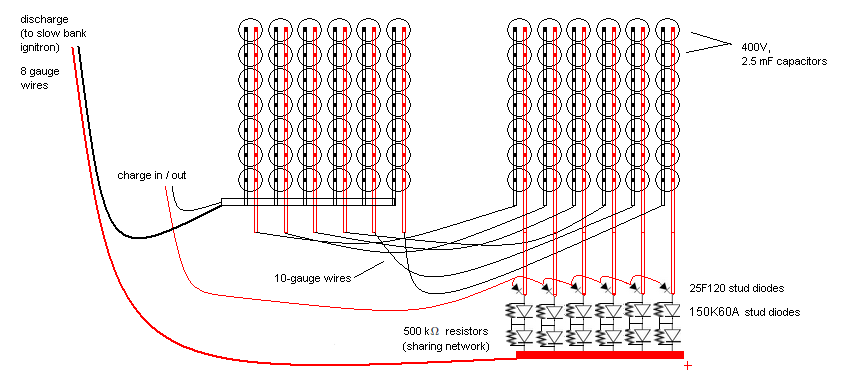}\caption{\label{fig:Slow-capacitor-bank1.}Slow capacitor bank \#1.}
\end{figure}
\begin{figure}[H]
\centering{}\includegraphics[scale=0.5]{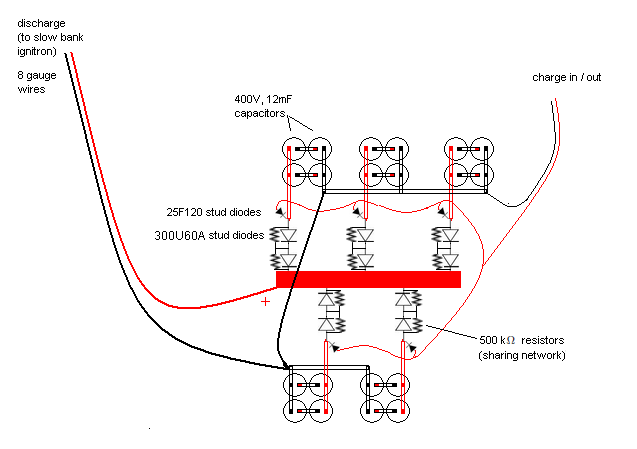}\caption{\label{fig:Slow-capacitor-bank2}Slow capacitor bank \#2}
\end{figure}
The original slow bank consisted of eighty-four 2.5mF, 450V capacitors
in parallel, with a total capacitance of 210mF at 450V and total stored
energy of 17kJ. Figures \ref{fig:Slow-capacitor-bank1.} and \ref{fig:Slow-capacitor-bank2}
shows the two sections of the extended slow capacitor bank. When they
are connected in parallel, the total bank capacitance is 112mF at
800V (36kJ). Bank 1 (800V, 52mF, 17kJ) consists of two 400V blocks
connected in series. Each block consists of six 400V sub-blocks connected
in parallel. Each sub-block is a row of seven 400V, 2.5 mF capacitors
connected in parallel. Charge current passes through one of six 25F120
diodes and charges the bank to 800V. When bank 1 is fully charged,
the voltage on the terminals of the individual capacitors in the left
block is 0V (-ve terminals) and 400V (positive terminals); in the
right bank the voltages are 400V (-ve terminals) and 800V (positive
terminals). The 25F120 diodes ($V_{RRM}$ = 1.2kV, $I_{F(AV)}$ (maximum
average forward current) = 25A) are there to prevent the capacitors
from discharging back along the charge lines. The 150K60A diodes ($V_{RRM}$
= 600V (so 1.2kV for two in series), $I_{FSM}$ = 3.7kA) are there
for safety; in the event that an individual capacitor fails, a short
circuit to the voltage of the negative terminal of the failed capacitor
would be presented to other capacitors connected to it. For example,
if a capacitor in the left block fails, each of the remaining 6 capacitors
(each charged to say 400V) in that row will discharge through the
failed capacitor to ground - this could cause an explosion. In addition,
current will flow through one of the 10-gauge wire connectors and
the failed capacitor to bring the voltage on the negative terminals
of each of the 7 capacitors in the connected row in the right block
from 400V to ground. Then the capacitors in the connected row in the
right block would be charged to twice their rated voltage and could
also explode. If the 150K60A diodes were not present, the energy stored
in the remaining 35 capacitors in the right block would also available
to contribute to the energy of the second explosion. On the other
hand, if one of the capacitors in a row in the right block failed,
the energy of either the 6 remaining capacitors in that row (diodes
present), or of the 41 remaining capacitors in the block (no diodes)
would be involved in a possible explosion. The $500\mbox{k}\Omega$
resistors are there for network sharing - to ensure that the voltage
being held off is equal across each of the two stud diodes in series;
if the voltage was shared unequally, one of the diodes could fail
if the voltage across it was over its rated maximum reverse voltage.
Two diodes in series are required since each of the diodes is rated
to 600V, while up to 800V must be held off.

Bank 2 (800V, 60mF, 19kJ) is composed of five blocks in parallel.
Each block consists of two sub-blocks in parallel, with each sub-block
being a pair of 400V, 12 mF capacitors in series. The diodes and resistors
present in the bank have the same functions as they do in bank 1.
The 300U60A stud diodes have ratings $V_{RRM}$ = 600V, $I_{FSM}$
= 6.7kA.

\subsubsection{\label{sub:Optically-triggered-6V}Optically triggered 6V pulse generators}

\begin{figure}[H]
\centering{}\includegraphics[scale=0.5]{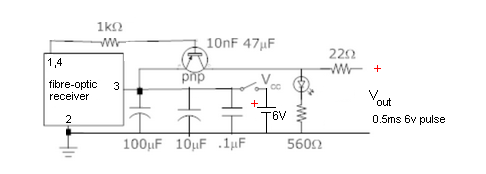}\caption{\label{fig:Optically-triggered-6V}Optically triggered 6V pulse generator
circuit.}
\end{figure}
Figure \ref{fig:Optically-triggered-6V} shows the circuit for the
optically triggered 6V pulse generators. When the unit is powered
on, the three capacitors are charged up to 6 volts. When it receives
an optical pulse, the fiber-optic receiver transmits a signal to the
pnp transistor base, allowing discharge of the capacitors through
the emitter and collector to form the required pulse and light an
indicator LED.

\subsubsection{\label{sub:Trigger-circuit-for}Trigger circuit for ignitron firing
control}

\begin{figure}[H]
\centering{}\includegraphics[scale=0.5]{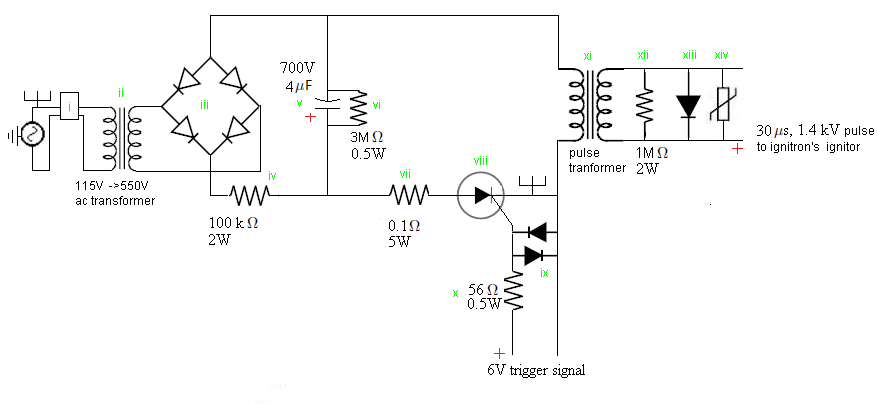}\caption{\label{fig:9}Ignitron trigger circuit\label{fig:Ignitron-trigger-circuit}}
\end{figure}
The trigger circuit for ignitron firing control is shown in figure
\ref{fig:Ignitron-trigger-circuit}. The principle function is to
respond to an incoming 6 Volt pulse generated by an optically triggered
circuit (see sub-section \ref{sub:Optically-triggered-6V}) by sending
a 1.4kV, $\sim30$$\upmu$s pulse to fire the ignitron. The 115V ac
supply is filtered (to remove voltage spikes that might damage other
circuit components ($e.g.,$ the bridge rectifier)) (i), then stepped
up to 550V at an ac transformer (ii). The output of the bridge rectifier
(iii) is a pulsating dc current at $~700$V. The $100\mbox{k}\Omega$
charge resistor (iv) is to limit current to within the limits set
by the transformer. A 700V, 4$\upmu$F capacitor (v) (with a $3\mbox{M}\Omega$
slow-drain resistor (vi) across it for safety) is charged. When the
SCR (viii) is turned on by the 6V trigger signal, the capacitor discharges
through a 1:2 step up transformer (xi) and the resulting 1.4kV pulse
to the ignitron's ignitor causes evaporation of mercury vapour in
the ignitron allowing the main (fast or slow) CHI capacitor banks
to discharge. Resistors (vii - ) and (x) limit current to protect
the SCR. The SCR (model 50RIA120, $V_{RRM}=1.2$kV) can handle maximum
trigger signal amplitudes of +1.5V and -0.5V; the diodes (ix) short
out signals above these limits. The function of diode (xiii) is to
short out any negatively biased voltage spikes (that may arise in
the circuit due to inductive back emf), preventing such spikes from
reaching the ignitor and possibly damaging the ignitron. Two metal
oxide varistors (P2T750E MOVS) in series were added to the system
and seem to have solved a problem with the overall main CHI discharge
circuit: as seen in figure \ref{fig:9-1}, the fast and slow capacitor
banks are arranged in parallel. A substantial part of the 5kV voltage
pulse appears transiently across the slow bank ignitron trigger electrodes
and travels to the slow bank trigger circuit. On countless occasions
this caused failure of the diode at (xiii) and/or failure of the SCR
(viii). Several investigations were undertaken to diagnose the cause
of the problem and find a solution. When the actual cause was unknown,
it was suspected that inductive pick up or over-current in the circuit
was causing the SCR failures. We experimented with EM shielding ($e.g.,$
putting the trigger circuit in a Faraday cage and distancing from
EM noise sources) and current limiting ($e.g.,$ increasing the resistance
at vii) methods. The slow bank ignitron and its trigger circuit were
replaced with an SCR and a diode stack (as in figure \ref{fig:Various-gun-current}(b).
Later that SCR failed and the ignitron was put back with the MOVs
in place. It proved to be an elusive solution but the MOV addition
seems to work. The MOV pair used shorts out voltage spikes that are
above the clamping voltage of 1.5kV, protecting the SCR.

\subsection{\label{sub:CHI-solenoid-circuit}CHI solenoid circuit}

\begin{figure}[H]
\begin{centering}
\includegraphics[scale=0.5]{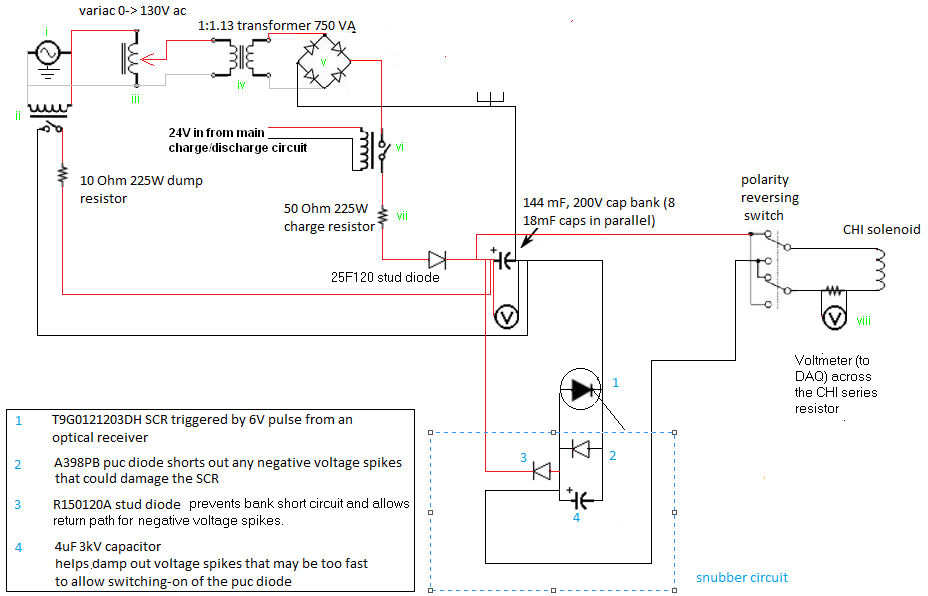}
\par\end{centering}

\caption{\label{fig:7}Solenoid charge/discharge circuit}
\end{figure}
The circuit for the charging and discharging of the solenoid capacitor
bank is shown in figure \ref{fig:7}. When the 115V alternating current
(ac) supply (i) is switched on, an electromagnet-powered relay (ii)
opens to prevent capacitor discharge through the 10$\Omega$ dump
225 Watt resistor. When the 2nd relay (vi) is closed by a 24V dc signal
(the signal is present during the \textquotedbl{}charge\textquotedbl{}
section of the cpf program if switch (11) and the relays (13) in figure
\ref{fig:9-1} are closed- see section \ref{sub:CHI-gun-current}),
ac voltage is stepped up or down with the variac (iii), stepped up
with a 750VA iron cored transformer (iv), rectified (and stepped up
by a factor of $\sqrt{2}$ to its rms value) to a pulsating (dc) state
at the bridge rectifier (v). Current flows through a 50$\Omega$ 225
Watt charge resistor (vii) and a 25F120 stud diode (this diode is
to prevent capacitor discharge through the bridge rectifier and transformer
coil - it has a $V_{RRM}$ (maximum repetitive peak reverse voltage
rating) of 1.2kV and an $I_{F(AV)}$ (maximum average forward current)
rating of 25 Amps) to charge the capacitor bank. The capacitor bank
is an array of eight 18mF 200V capacitors arranged in parallel to
form a 200V bank with a capacitance of 144mF. The discharge circuit
is kept open until the T9G0121203DH puck SCR (silicon controlled rectifier-
$V_{RRM}$ (maximum repetitive peak reverse voltage rating) = 1.2kV,
$I_{FSM}$ (max. surge current) $=$27 kA) is turned on by a 6V pulse
from an optically-triggered pulse generator (see section \ref{sub:Optically-triggered-6V}).
When the SCR is triggered, the capacitor bank discharges through the
solenoid polarity-reversing-switch, the CHI solenoid, the CHI solenoid
series resistor and the SCR. The snubber circuit indicated in figure
\ref{fig:7} acts to prevent damage to the SCR that could arise due
to high amplitude ($i.e.,$ above 1200V) (short duration) negative
voltage spikes that can be inductively induced during switching of
the discharge circuit. The functions of the snubber circuit components
are indicated in the figure. 

The polarity reversing switch allows CHI device (envisaged) operation
with production of plasmoids with both positive and negative helicity
(with fixed polarity on the gun electrodes). The reading from the
voltmeter (a differential probe) across the solenoid series resistor
(viii) is sent to the DAQ (data acquisition) system and the current
and stuffing field can be calculated and plotted from that data. The
original resistor (1$\Omega$, 100W) can be replaced with various
combinations of resistors to vary the magnitude of the solenoid current
(and stuffing field). For example a 0.2$\Omega$ (0.01$\Omega$ -
$i.e.,$ 10' of 10-gauge wire) resistor increases the magnitudes by
a factor of five (100) with regard to  the case with a 1$\Omega$
resistor. The inductance of the CHI solenoid can be calculated with
the formula for a long (length > radius) solenoid: $L=\mu N^{2}A/l$,
where $\mu$ is the magnetic permeability of the solenoid core material
(in this case the core is air so $\mu\approx\mu_{0}=4\pi\times10^{-7}$),
N is the number of solenoid-wire turns, $A=\pi r^{2}$ where r is
the solenoid radius and $l$ is the solenoid's length, giving a value
of $L\approx17.5$nH. Modification of the original circuit during
this project included the replacement of the original 0.1Amp charging
system with a variable charging system rated to a maximum of 4 Amps.
This allowed faster charging of the bank to full capacity and also
involved replacing components on the charge side of the circuit to
be able to handle the larger current. Also, the relay system with
the dump resistor was added for safety and the polarity reversing
switch was added.

\section{\label{sec:Results-of-CHI}Results of CHI injection to STOR-M}

\begin{figure}[H]
\centering{}\subfloat[]{\includegraphics[scale=0.4]{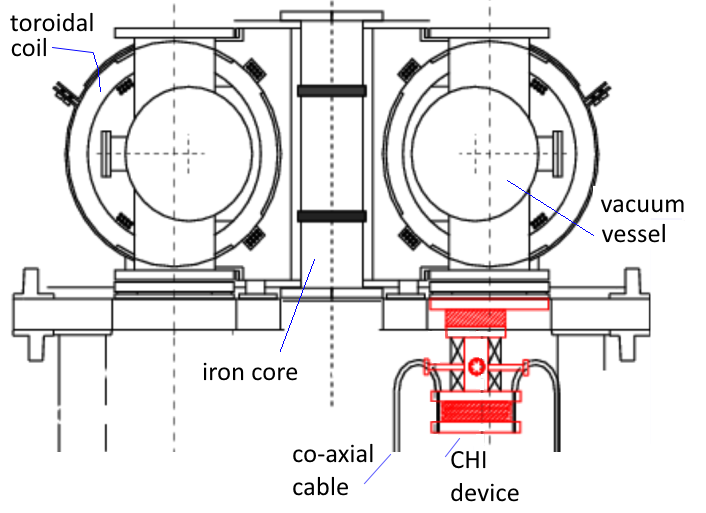}

}\hfill{}\subfloat[]{\includegraphics[scale=0.4]{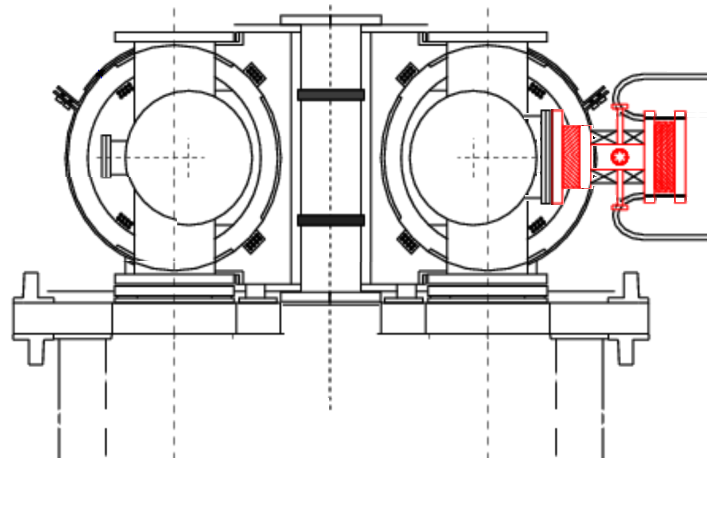}

}\caption{\label{fig:Vertical-CHI-mounting}CHI mounting positions}
\end{figure}
The CHI device was initially mounted on the STOR-M tokamak as shown
in figure \ref{fig:Vertical-CHI-mounting}(a). Injection to the STOR-M
plasma using the original fast (2$\upmu$F,$\:5$kV) and slow ($210\mbox{mF},\:450$V)
banks did not show any indication of an increase in tokamak toroidal
current ($I_{p}$) (as measured by a Rogowski coil on the vacuum chamber).
The usual effect on tokamak plasma was a disruption accompanied by
a rise in loop voltage. When the slow bank was modified to include
banks 1 and 2 ($112\mbox{mF},$$\:800$V - see section \ref{sub:Slow-capacitor-bank.}),
increases in $I_{p}$ of around 3-5kA over 3-5ms were routinely observed.
In this configuration the slow bank circuit was modified to allow
for a negatively biased central CHI device electrode and the fast
bank had to be disconnected, as described at the end of section \ref{sub:CHI-gun-current}.

\begin{figure}[H]
\begin{centering}
\includegraphics[scale=0.5]{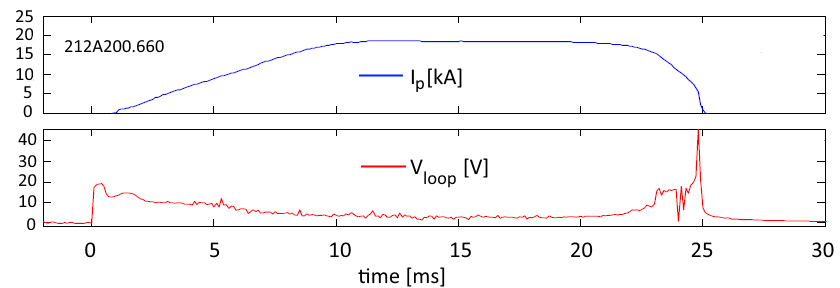}
\par\end{centering}

\caption{\label{fig:Reference-STOR-M-shot}Reference STOR-M shot without CHI
injection}
\end{figure}
The indicated plasma current increase was usually followed by disruption
of the plasma. An investigation involving variation of parameters
including fast bank voltage, slow bank voltage, solenoid voltage,
gas puff duration, gas puff voltage, gas puff timing, tokamak gas
puffs and tokamak vertical positioning was undertaken with the objective
of maintaining the current rise while mitigating the disruption. Figure
\ref{fig:Reference-STOR-M-shot} shows the plasma current and loop
voltage for a typical STOR-M shot without CHI injection, with a plasma
duration of around 25ms, a peak $I_{p}$ of around 20kA and loop voltage
of around 3 to 4 Volts during the current flat-top.

\begin{figure}[H]
\begin{centering}
\includegraphics[scale=0.5]{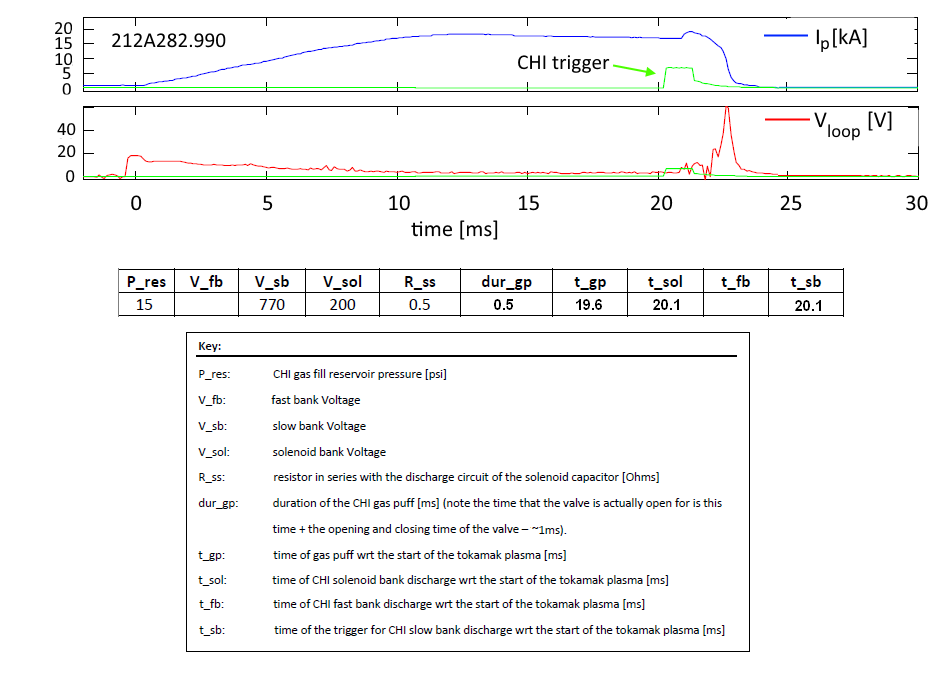}
\par\end{centering}

\caption{\label{fig:Shot-262.990}Shot \#262.990}
\end{figure}
Plasma current and loop voltage are shown for a typical STOR-M discharge
with CHI injection (no fast bank) in figure \ref{fig:Shot-262.990}.
It was found that the fast bank was not required for CHI plasma formation
for slow bank voltages above around 350V. The time (an input to the
STOR-M cfp program) of the optical pulse (2ms duration here - actually
duration of this signal is not important as long as it is above the
threshold of the fiber-optic receiver described in section \ref{sub:Optically-triggered-6V})
to trigger the slow bank discharge is indicated by the green trace.
The main CHI system parameters for the shot, and an explanation of
the parameters, are included in the figure. Note the (indicated) 4kA
rise in plasma current, followed by disruption, at around 1ms after
the trigger.

\begin{figure}[H]
\centering{}\includegraphics[scale=0.5]{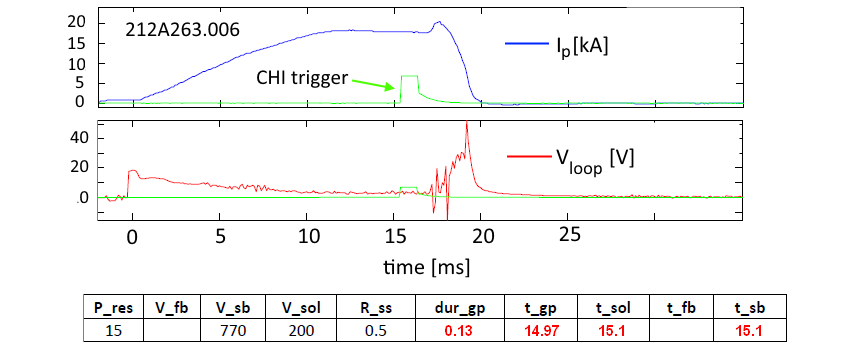}\caption{\label{263.005}Typical shot with CHI injection}
\end{figure}
The gas puff duration (in red) was decreased, and the CHI plasma was
injected earlier (compared with shot \#282.990), in shot \#263.006
with results shown in figure \ref{263.005}. It was thought that the
disruption following the current rise may have been caused by the
injection of cool neutral gas entrained with the CHI plasmoid, so
it is desirable to keep the gas puff duration to the minimum possible
for breakdown. Shorter gas puffs were found to result in a longer
delay between the slow bank trigger and formation of the CHI plasma.
Note the solenoid polarity in \#282.990 was reversed with regard to
that in \#263.006 - there is no obvious major effect. Many shots were
taken with all parameters besides solenoid polarity equal, and it
was found that polarity switching had no observed effect. Solenoid
polarity reversal should result in the formation of a spheromak with
a change in sign of helicity. Positive helicity is defined \cite{Brown_Bellan}
with the right hand rule; with the thumb pointing in the direction
of the spheromak's toroidal field, the fingers curl around the thumb
in the direction of the spheromak's poloidal field. According to the
theory of helicity injection as a tokamak plasma current drive mechanism,
the signs of the helicity content of the tokamak and injected spheromak
should match for increased plasma current and vice versa \cite{Brown_Bellan}.
If the current modification apparent with CHI injection was truly
a result of helicity injection, then it would be expected that two
injections with opposing helicity sign would have lead to a current
increase in one shot and a decrease in the other. This was an early
indication that the current modification signals obtained were either
spurious or at least not the result of helicity injection.

In most cases, gas was injected with a duration such that the end
of the gas puff (time at which the piezo-electric valve begins to
close) coincided with the trigger for the CHI slow bank. Overlapping
the end of the puff with the slow bank trigger was found to have no
effect on the duration of the slow bank discharge. It was found that
the minimum gas puff duration required for CHI plasma breakdown decreased
with increasing stuffing field until a point, and then increased with
the stuffing field.

Several shots were done to show the effect of CHI injection without
tokamak plasma, but with varying $B_{T}$ (tokamak toroidal field).
$B_{T}$ seemed to have no effect on the $I_{p}$ signals observed.
The fact that the $I_{p}$ is seen to arise (max 5kV over $~4$ms)
even with no $B_{T}$ and that the $B_{T}$ strength has apparently
no effect on the $I_{p}$ signal indicated that perhaps no real plasma
current was being driven. Without a toroidal field, confinement of
the CHI plasma particles, and therefore a toroidal current, should
be impossible. 

The CHI device was mounted radially (see figure \ref{fig:Vertical-CHI-mounting}(b))
because it was thought that more effective helicity injection could
be acheived with a shorter discharge tube - the fields of the injected
plasmoid would have less time to decay before merging with the tokamak
plasma. The discharge circuit was changed again to have a positively
biased central electrode. No major change in the effect of injection,
with regard to injections with the vertical mounting, on tokamak plasma
current and loop voltage was observed. However, the duration of the
$I_{p}$ spikes decreased in general, and it was found that the gas
puff requirements for CHI plasma breakdown increased significantly.
The minimum gas puff duration at 15psi (gauge) reservoir pressure
increased from 0.1 to 2 ms in the radial mounting position. This may
have been because the piezo-electric valve was not opening as well
as it used to, or possibly due to the effect of the two 30cm lengths
of corrugated half-inch diameter vacuum pipe that injected gas must
travel through to reach the CHI device. Previously the gas from the
valve travelled around 10cm to a single entry point on the CHI device;
in the newer configuration (motivated by a desire to get a more radially
uniform gas delivery and plasma breakdown) the path of the gas was
split in two to reach the device through two entries located on opposite
sides. The original 5kV, $2\upmu$F fast bank was intended as an aid
to assist with CHI plasma breakdown. However it was found that plasma
could be formed with slow bank voltages above around 340V. Inclusion
of fast bank operation allowed for operation at lower slow bank voltages
(around 230V) and allowed for slightly lower gas puff durations on
some plasma-producing shots, but seemed to have no effects on others.
A 0.5mF fast bank was added to the discharge circuit (figure \ref{fig:Various-gun-current}(d))
not for breakdown assistance, but rather in the hope that more acceleration
of the plasmoid would result in actual helicity injection. 
\begin{figure}[H]
\centering{}\includegraphics[scale=0.5]{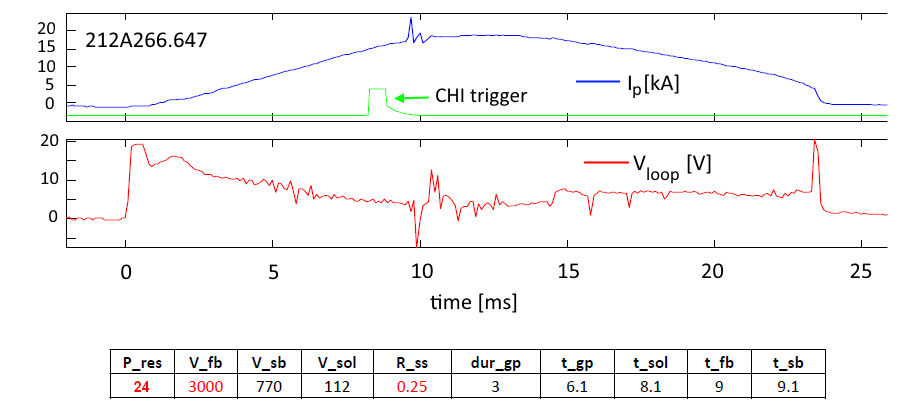}\caption{\label{fig:Shot-266.647-Radial}Shot 266.547 Radial mounting with
0.5mF fast bank}
\end{figure}
Figure \ref{fig:Shot-266.647-Radial} shows the signals obtained for
a shot with radial mounting, with 3kV on the 0.5mF fast bank. Injection
here did not cause an immediate distruption of tokamak plasma. Also
indicated in the figure are the gun voltage, gun current and solenoid
current signals.

It was found that CHI operation induced fluctuations on the $I_{p}$
and $B_{T}$ signals, with approximately the same ampltiude, even
without STOR-M plasma and $B_{T}.$ To investigate further, shots
were taken with the $B_{T}$ ignitron trigger circuit powered off
and the $B_{T}$ fluctuation was still induced despite having an open
$B_{T}$ circuit. This eliminated the possibility that an effect of
injection was to induce current flow in the $B_{T}$ coils. Signals
similar to the spurious $B_{T}$ signals were also found to appear
on other channels that have Rogowski coil induced voltage signals
as their source. Several CHI injections (without tokamak plasma) were
done with the $I_{P}$ Rogowski coil removed from its usual position
around the tokamak vacuum chamber and placed in various temorary positions
at various orientations in the same general vicinity as its usual
position. This test to verify that the $I_{P}$ signals were indeed
spurious was not conclusive though the apparent current increases
did vanish unless the coil enclosed the chamber. A series of definitive
tests showed finally that the apparent increase in $I_{p}$ with injection
was indeed due to spurious signals. The CHI device was disconnected
from STOR-M, vacuum sealed, and replaced in approximately the same
location by its radial mounting port. It was connected to a portable
vacuum pump and several CHI discharges were done without tokamak plasma.
The tokamak $I_{p}$ signal continued indicated a current rise of
several kiloamps. For some of these shots the flange at the CHI device
exit was connected by a conductor to the tokamak vessel. Current was
observed to flow to the tokamak chamber along the wire. Although this
current was not measured, it was significant enough to burn out a
16 gauge wire over a discharge of order 10ms. The $I_{p}$ signals
observed with this connection tended to have peaks comparable to those
without connection. This seems to indicate that the signals picked
up by the $I_{p}$ Rogowski coil was largely inductively rather than
capacitatively induced.

\section{\label{sec:CHI-Plasma-Characterisation}CHI Plasma Characterisation}

It was decided to characterise the plasmoid produced by the CHI device.
A main point of the investigation was to see if the plasmoid being
produced was actually an \textquotedbl{}elongated spheromak\textquotedbl{},
or perhaps a series of spheromaks, as had been envisaged. A \textquotedbl{}T\textquotedbl{}
 junction pipe section with Con Flat vacuum flanges and five available
ports for diagnostics with an internal diameter of around 6\textquotedbl{}
was available for modification to be used as a test chamber. The \textquotedbl{}bottom
of the T\textquotedbl{} was blocked off with a sheet of rolled 2mm
thick stainless steel to preserve magnetic flux conservation at the
CHI plasmoid boundary. To maintain a spheromak-type plasmoid, the
field should remain parallel to the boundary in a cylindrical flux
conserver. The magnetic diffusion time can be calculated from the
formula $\tau_{diff}=t^{2}\mu/\eta$. For 304 stainless steel, $\mu_{r}=\mu/\mu_{0}\approx1$,
and $\eta\approx1.16\times10^{-6}\,\Omega-$m. The thickness ($t$)
of the walls of the chamber itself is around 4mm so $\tau_{diff}\sim17\upmu$s.
For the section with the 2mm thick wall, $\tau_{diff}\sim4\upmu$s.
Using the plasmoid speed measured on a similar co-axial plasma injector
of 17km/s \cite{Asai1} as an estimate, it can be estimated that the
plasmoid will require $\sim19\upmu$s to travel the 35cm distance
to the downstream probe, so lack of flux conservation may be an issue.
Magnetic probes were constructed, to measure poloidal and toroidal
field near the CT edge, and a Helmholtz coil was used to calibrate
the probes. Measurements were taken of the $r$, $\phi$, and $z$
magnetic field components, at two locations, one near the CHI device
exit and the other around 30cm downstream. Langmuir probes were made
and returned reasonable estimates for edge CT density and temperature.
\begin{figure}[H]
\centering{}\includegraphics[scale=0.5]{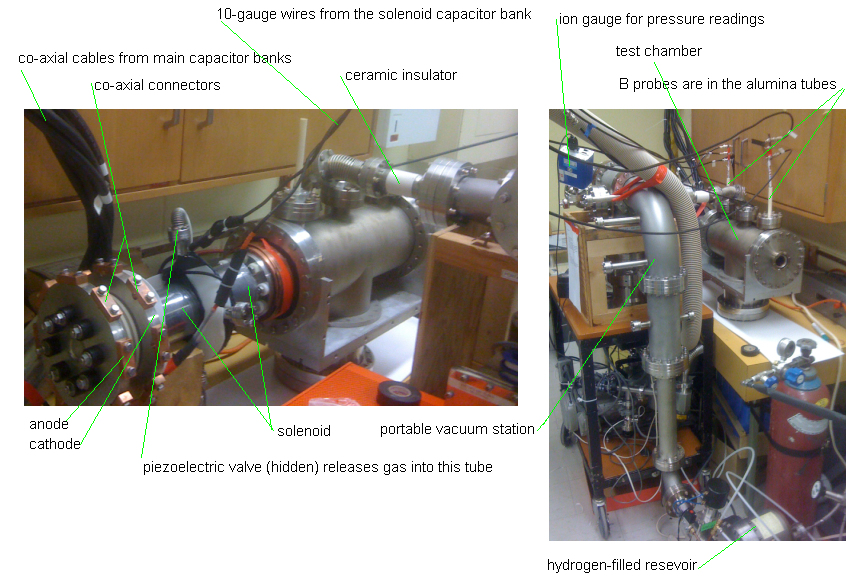}\caption{\label{fig:CHI-device-on}CHI device on test chamber}
\end{figure}
Figure \ref{fig:CHI-device-on} shows the experimental set-up for
CHI plasma characterisation, with the main components indicated. The
pressure attained in the chamber was of the order of $10^{-6}$ Torr.
The hydrogen reservoir was filled to around one atmosphere (760 Torr). 

Poloidal and toroidal magnetic fields measured at the magnetic probe
located by the gun exit, with the probe inserted around 2cm into the
test chamber, were typically around 0.05T and 0.1T respectively. The
signals were very attenuated at the location of the magnetic probe
located around 30cm downstream of the gun exit, perhaps due to lack
of adequate flux conservation. Hence, magnetic probe data could not
be used to estimate the time of flight of the plasmoid. Langmuir probe
analysis indicated edge density and temperature of around 5$\times$10$^{18}${[}m$^{-3}${]}
and 10eV. It may be the case that the sudden expansion of the plasmoid
at the CHI device exit (diameter around 6.5cm) when it enters the
chamber (diameter 15cm) disrupts the plasmoid structure. Ideally,
an expansion cone to allow a gradual flux-conserving transition to
the chamber would have been installed. The chamber itself had a diameter
similar to that of the radial port used for injection on STOR-M so
it is suitable in that sense. On the other hand, it may be better
to test the plasmoid in a chamber of diameter similar to that of the
CHI device exit and consider injection to STOR-M through a port of
similar size.

\section{Summary\label{sec:SummaryCHI}}

Although the tests conducted indicated that the CHI device was not
successful in injecting helicity to STOR-M, most of the difficulties
encountered with the various circuits that control its operation were
overcome, and its operation under various combinations of conditions
was assessed. It would be necessary to characterise the plasmoid produced
by the CHI device and, if possible modify it to produce verified spheromaks,
before attempting injection on STOR-M again. The geometry of the test
chamber may need to be modified. A spectroscopic investigation should
be undertaken to see what impurities are present in the plasmoid.
The cathode should be tungsten coated to reduce plasma impurities
that can lead to radiative cooling. As mentioned in section \ref{sec:Results-of-CHI},
the gas injection system was modified to allow more uniform injection
through two ports instead of one, and that it is suspected that this
arrangement with two 30cm long flexible vacuum pipes has had the effect
of delaying the delivery of gas. A more efficient and uniform gas
delivery method may be to bore a hole axially through the central
electrode from the plate (see figure \ref{fig:3}) as far as the port
region, and bore a further set of several holes with smaller-radius
radially inwards to meet the axial bore. The gas valve could then
be attached with a ceramic insulator pipe to the back of the central
anode plate. 

If spheromak production is confirmed, injection to STOR-M should possibly
be done with the tokamak chamber at a potential other than the CHI
cathode potential. The CHI device was mounted on the hard-grounded
side of STOR-M's vacuum chamber - the other side is soft-grounded
through $10\mbox{k}\Omega$. In the proposed CHI operation scenario,
current should flow initially in a radial direction from the anode
to cathode, then should flow along the stuffing field from anode to
cathode. It may be that current returned to ground via the tokamak's
ground instead of via the CHI cathode.

\section{Acknowledgments}

Funding was provided in part by the University of Saskatchewan and
NSERC. We would like to thank to Dave McColl, Tomohiko Asai and Masaru
Nakajima for technical assistance and useful discussions.

\end{document}